\documentclass{aastex631}

\usepackage{booktabs}
\usepackage{amsmath}
\usepackage{subfigure}
\usepackage{cancel}
\begin{document}

\title{Early $\rm TeV$ photons of GRB 221009A were absorbed by the prompt MeV photons}

\correspondingauthor{Yuan-Chuan Zou}
\email{zouyc@hust.edu.cn}

\author[0009-0007-0866-4265]{Duan-Yuan Gao}
\affiliation{School of Physics, Huazhong University of Science and Technology, Wuhan 430074, China}

\author[0000-0002-5400-3261]{Yuan-Chuan Zou}
\affiliation{School of Physics, Huazhong University of Science and Technology, Wuhan 430074, China}



\begin{abstract}
Gamma-ray burst (GRB) 221009A produced the highest flux of gigaelectronvolt-teraelectronvolt ($\rm GeV-TeV$) photons ever observed, allowing the construction of a detailed $\rm TeV$ light curve. We focus on explaining the noticeable dip in the light curve around $2$-$5\ \rm s$ after the onset of $\rm TeV$ emission. We propose that Megaelectronvolt (MeV) photons from the prompt emission annihilate with $\rm TeV$ photons from the afterglow, producing an optical depth that obscures the $\rm TeV$ emission during this period. We develop a two-zone model accounting for the angles of MeV photons that can successfully reproduce the time delay between MeV and $\rm TeV$ photons, the peak optical depth over 3, and the rapid decline in optical depth. Our model supports $\rm MeV-TeV$ annihilation as the cause of the dip and provides reasonable constraints on the emission region parameters.
\end{abstract}



\section{Introduction} \label{sec:intro}
%
GRB 221009A stands out as the brightest gamma-ray burst (GRB), highlighting its extreme rarity with an occurrence rate of approximately once every ten thousand years \citep{2023ApJ...946L..31B}. It has a relatively low redshift of $z=0.151$ \citep{2022GCN.32648....1D}. The trigger time ($T_{\rm 0}$) was recorded on October 9, 2022, at 13:16:59.99 UTC \citep{2023arXiv230314172L}. The Fermi Gamma-ray Burst Monitor (GBM) observed its emission across the energy range of 8 keV to 40 MeV \citep{2023arXiv230314172L}. Furthermore, Insight-HXMT and GECAM-C contributed observations within the energy span of 10 keV to 6 MeV \citep{2023arXiv230301203A}. A joint observation effort was also conducted, involving Konus-Wind in the energy range of 20 keV to 1.22 MeV and ART-XC spanning 4-120 keV \citep{2023ApJ...949L...7F}.

Around $226\ \rm s$ after the trigger, the Large High Altitude Air Shower Observatory (LHAASO) detected a remarkable count of 64,000 $\rm TeV$ photons \citep{2023Sci...380.1390.}. 
The presence of $\rm TeV$ photons in the afterglows of preceding gamma-ray bursts has been documented as well: GRB 180720B exhibited photons with energies ranging from 100 to 440 GeV \citep{2019Natur.575..464A}. GRB 190114C showed photons with energies spanning 0.2 to 1 $\rm TeV$ \citep{2019Natur.575..455M}. Similarly, in GRB 190829A, photons with energies between 0.18 and 3.3 $\rm TeV$ were also identified \citep{2020ApJ...898...42C}.

The notably higher flux of $\rm TeV$ photons from GRB 221009A facilitated the detection of a more substantial photon count, enabling the assembly of a $\rm TeV$ photon light curve. While the majority of this light curve aligns well with the afterglow model, a discernible dip emerges in the light curve approximately $2$-$5\ \rm s$ after the onset of $\rm TeV$ photons \citep{2023Sci...380.1390.}. 
Through fitting, they found a rapid rise in the early stage of TeV radiation that cannot be explained by the afterglow model.

$\rm TeV$ photon emission of this GRB has been studied in several works.
Some researchers suggested that the $\rm TeV$ photons originate from the prompt emission. \cite{2023SCPMA..6689511W} explored the possibility of the prompt origin of TeV photons and found it is plausible that the hadronic process in the prompt regime can contribute to TeV emission.
From the smoothness of the $\rm TeV$ light curve  \citep{2023Sci...380.1390.}, one may conclude that it comes from the external shock.
\cite{2023arXiv231011821W, 2023ApJ...956...12I} modeled the whole multiband light curve from $\rm keV$ to $\rm TeV$, and they all found both the hadronic and leptonic processes should have contributed to the multiband emission, while \citet{2023arXiv231015886R} suggested a structure jet which can explain the multiband emission.
Previous studies have also investigated the optical depth of $\rm TeV$ photons in GRB 221009A. \cite{2023arXiv230714113D} examined the optical depth of photon annihilation under their Internal-Collision-Induced Magnetic Reconnection and Turbulence (ICMART) scenario, which is insufficient to block $\rm TeV$ photons.
In \cite{2023arXiv231000631G}, we studied the optical depth generated by the annihilation of $\rm TeV$ photons with MeV photons. However, at that time, we were restricting the minimum Lorentz factor. We only considered the time period after $5\ \rm s$ when it was certain that $\rm TeV$ photons were not obscured.
\cite{2023arXiv230810477S} considered various mechanisms for generating optical depth, including: $\rm TeV$ photons generating electron-positron pairs through cascading, leading to scattering; absorption of $\rm TeV$ photons by electrons in the external shock; absorption of $\rm TeV$ photons by electrons in the ejecta; and annihilation of low-energy photons produced by the external shock with $\rm TeV$ photons.

If we extend the second segment of the LHAASO fitting curve to earlier times, we can find that it nearly perfectly passes through the point
, just before the dip. Combining this with the light curve of MeV photons, we can find that the time of the dip is close to the time when the MeV brightness starts to rise. Therefore, we hypothesize that the MeV photons from the prompt burst and the $\rm TeV$ photons from the afterglow annihilate each other \citep[two-zone model, ][]{2011ApJ...726L...2Z}, blocking the $\rm TeV$ photons during that period. As the distance increases, MeV photons produce insufficient optical depth, allowing subsequent $\rm TeV$ photons to pass through. In section \ref{sec:tw}, we will use two different approximations of the two-zone model that considers the annihilation of MeV and $\rm TeV$ photons to explain the dip in the light curve. In section \ref{sec:con}, we will summarize the findings of the paper and provide prospects.

\section{The two-zone model used to explain the dip}
\label{sec:tw}
We assume that the second segment of the LHAASO fitting curve illustrates the light curve before being obscured by MeV photons. 
We deduce the optical depths and their corresponding error margins at different data points, which are shown as the 
orange data points in the lower panel of Figure \ref{fig:mainresult}. 
Upon examination, we identify three obstacles in {explaining} the dip within the light curve: first, {explaining} the temporal delay between the upsurge of MeV photons and the {beginning} of the dip;
second, attaining a peak optical depth exceeding 3; and third, clarifying the swift decline in optical depth. 
Subsequently, we will strive to address these challenges through the utilization of varied approximations of the two-zone model.
In the two-zone model, $\rm TeV$ photons and MeV photons originate from different emission regions, and the annihilation of MeV photons prevents the proceeding of $\rm TeV$ photons.
The MeV photons come from an {emission} region with a radius $R_{\rm{M}}$, a Lorentz factor of $\Gamma_{\rm{M}}$, and an angular width of $\theta_{\rm{M,J}}$. 
Similarly, the $\rm TeV$ photons come from a different {emission} region with a radius $R_{\rm{T}}$, a Lorentz factor of $\Gamma_{\rm{T}}$, and an angular width of $\theta_{\rm{T,J}}$.

\subsection{The case dominated by the size of the $\rm TeV$ photon emission region}
In this scenario, we assume that the emission radius of $\rm TeV$ photons is much larger than that of MeV photons, thus neglecting the impact of the size of the MeV emission region on the collision angles between photons.
Similar to \cite{2023arXiv231000631G}, we employ the concept of {retarded} time $t_{\rm R}$ to characterize the timing of the light curve. 
We take the trigger time and the Earth's position as the zero point, where the retarded time equals the time on the light curve.
Furthermore, we assume that $R_{\rm M}$ is significantly smaller than $R_{\rm T}$, resulting in a time delay between $\rm TeV$ and $\rm MeV$ photons $\delta t_{\rm{R}}$, as illustrated below:
\begin{equation}
 \delta t_{\rm{R}} = t_{\rm R,T} - t_{\rm R,M}  \approx \frac{1}{2}(1+z)\theta^2 \frac{R}{c},
  \label{eq:tm}
\end{equation}
where $t_{\rm R,T}$ is retarded time of $\rm TeV$ photons, 
 $t_{\rm R,M}$ is retarded time of MeV photons, $c$ is the speed of light, $z$ is the redshift, $R$ represents the radius to the central engine and $\theta$ is collide angle between $\rm TeV$ photons and MeV photons. 

The collision angle between $\rm TeV$ photons and MeV photons can be expressed as:
\begin{equation}
    R \sin\theta = R_{\rm{T}} \sin\theta_{\rm{T}} = {\rm const.} \approx R \theta. 
\end{equation}

We can employ the Band function \citep{1993ApJ...413..281B} to characterize the spectrum, incorporating parameters such as the number density $n_{0,\rm{MeV}}$, the power-law photon indices $\alpha_{\rm M}$ and $\beta_{\rm M}$, and peak energy $E_{\rm{P}}$, all of which are time-dependent:
\begin{eqnarray}
 n_{\rm{MeV}}(E_{\rm{M}}) \simeq n_{\rm{0},\rm{MeV}} (R,t_{\rm{R}})  \left\{  
  \begin{array}{ll}
   \left[\frac{E_{\rm{M}}}{E_{\rm{P}}(t_{\rm{R}})}\right]^{-\alpha_{\rm{M}}(t_{\rm{R}})}, \phantom{00} & E<E_{\rm{P}}(t_{\rm{R}}) , \\
   \left[\frac{E_{\rm{M}}}{E_{\rm{P}}(t_{\rm{R}})}\right]^{-\beta_{\rm{M}}(t_{\rm{R}})}, \phantom{00} & E_{\rm{p}}(t_{\rm{R}})<  E < E_{{\rm{M}},{\max}}.
  \end{array}
   \right. 
\end{eqnarray} 
Since the observed photon energy lower limit is much lower than the peak energy, we only consider the upper limit of integration. The number density of MeV photons per unit energy is:
\begin{equation}
    n_{0,\rm{Mev}} = \frac{L_{\rm{M}}(t_{\rm{R}})}{(1+z)^2 4\pi R^2c \int_{0}^{E_{\max,\rm{o}}} E \left(\frac{E}{E_{\rm{P}}(t_{\rm{R}})}\right)^{-{\alpha}} {\rm d}E } = \frac{L_{\rm{M}}(t_{\rm{R}})}{(1+z)^2 4\pi R^2c {E_{\rm{P}}^{*2}(t_{\rm{R}})}},
    \label{eq:n0}
\end{equation}
where $E_{\max,\rm{o}}$ represents the maximum energy of the observed MeV photon and $L_{\rm{M}}$ represents the isotropic luminosity of MeV photons. The power-law photon index $\alpha$ can be either $\alpha_{\rm{M}}$ or $\beta_{\rm{M}}$ depending on the specific energy range. {$E_{\rm P}^*$ can be found in our previous article \citep{2023arXiv231000631G}.}

The scattering cross section $\sigma$, the minimum energy required for the MeV photons to annihilate with the $\rm TeV$ photons $E_{\rm M,\min}$ and the maximum value of the radius can be obtained from \cite{2023arXiv231000631G}. By combining the aforementioned equations, we can compute the optical depth of $\rm TeV$ photons as follows:
\begin{equation}
\tau(\theta_{\rm{T}},E_{\rm{T}}) = \int_{R_{\rm{T}}}^{R_{\max}(R_{\rm{T}},\theta_{\rm{T}})} {\rm d}R  \int_{E_{\rm M,\min}}^{E_{\rm M,\max}} {\rm d}E_{\rm{M}}  \frac{{\rm d} ^3 n_{\rm{MeV}}(t_{\rm{R}})}{{\rm d}E_{\rm{M}} {\rm d}\theta_{\rm{M}}} \sigma (1-\cos \theta),
\label{eq:tau_theta}
\end{equation}
where $E_{\rm M,\max}$ denotes the upper limit of the MeV component.
The effective optical depth, averaged over all angles, can be expressed as:
\begin{equation}
e^{-\bar{\tau} (E_{\rm{T}})} = \frac{\int_{0}^{\theta_{\rm T,j}}  \mathcal{D}^{-(3+\beta_{\rm{T}})} e^{-\tau(\theta_{\rm{T}},E_{\rm{T}})}     \theta_{\rm{T}} {\rm d}\theta_{\rm{T}}}{ \int_{0}^{\theta_{\rm T,j}} \mathcal{D} ^{-(3+\beta_{\rm{T}})}    \theta_{\rm{T}} {\rm d}\theta_{\rm{T}}},
\label{eq:e_tau_complete}
\end{equation}
where $\mathcal{D}=\Gamma_{\rm{T}}({1-\beta_{\rm bulk,T} \cos \theta_{\rm{T}}})$ is the Doppler factor, $\beta_{\rm bulk,T}=(1-1/\Gamma_{\rm{T}}^2)$ is the bulk velocity in $c$, and $\beta_{\rm{T}}$ is the photon index of the $\rm TeV$ emission. 
In our case, due to the beaming effect, $\theta_{\rm T} = 1/\Gamma_{\rm T}$.

When attempting to explain the $\rm TeV$ photon light curve, this model encountered two challenges. 
One problem is the inability to simultaneously achieve a large optical depth and a significant time delay. This {arises} from the fact that a high optical depth necessitates a sizable collision angle and a shorter distance. In this model, the collision angle is correlated with the Lorentz factor of the $\rm TeV$ photons, indicating that a lower Lorentz factor is required for the $\rm TeV$ photons. However, a substantial time delay requires a larger radius, implying a higher Lorentz factor.
Another issue involves the inability of the optical depth to decrease rapidly, which will be elaborated upon through a detailed comparison later. This characteristic can also be deduced from the formula of the simple model \citep{2011ApJ...726L...2Z}:
\begin{equation}
     \tau(\theta_{\rm{T}}) \approx \frac{(1+z)^{2{\alpha}-4}L_{\rm{M}}\epsilon_{\rm{P}}^{\alpha-2} \epsilon_{\rm{T}}^{{\alpha}-1} \sigma_{\rm{T}}\theta_{\rm{T}}^{2{\alpha}}}{6\times 4^{\alpha}\pi R_{\rm{T}} m_{\rm{e}}c^3{\alpha}},      
    \label{eq:simple_tua}
\end{equation}
where $\alpha=\alpha_{\rm M}$, $m_{\rm e}$ is the rest mass of an electron, $\epsilon_{\rm{P}} \equiv E_{\rm{P}}/m_{\rm{e}}c^2$, $\epsilon_{\rm{T}}\equiv E_{\rm{T}}/m_{\rm{e}}c^2$, $\sigma_{\rm T}$ is the Thompson cross section. We ignore the effect of {the} light curve.
In this equation, only $R_{\rm T}$ is related to $t_{\rm T}$, so $\tau \propto 1/t_{\rm T}$, which decays very slowly. 
Therefore, it {cannot} explain the dip.

\subsection{The case dominated by the size of the MeV photon emission region}

\begin{figure}[ht!]
\plotone{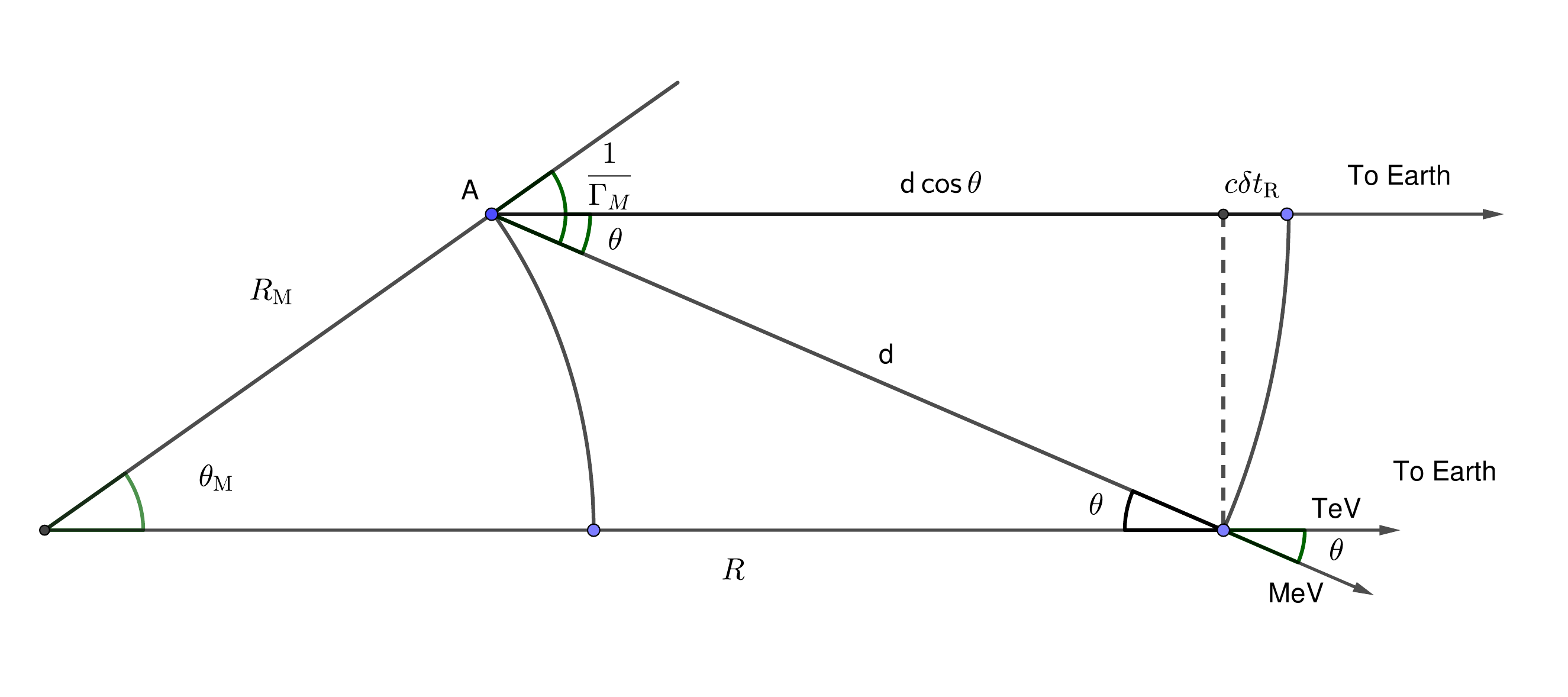}
\caption{A schematic diagram illustrates the {difference in the} time delay between MeV and $\rm TeV$ photons caused by geometric factors, as well as the correlation among various angles.
We assume that MeV photons emanate from the widest angle, denoted as $\theta_{\rm M}$, where the angle between the radial direction and the photon's trajectory is also maximized, at $1/\Gamma_{\rm M}$.
Conversely, we {assume} that $\rm TeV$ photons traverse the path that extends from the central engine to Earth, implying $\theta_{\rm T} = 0$.
Consequently, the collision angle between the $\rm TeV$ and MeV photons is $\theta$. 
It is important to note that the MeV photons colliding with the $\rm TeV$ photons do not reach Earth. However, we can calculate their particle density using photons arriving at Earth {at} the same retarded time.
In the illustration, the $\rm TeV$ and MeV photons share the same temporal moment, resulting in the retarded time difference being contingent upon the distance. MeV photons originate from point A; hence their { retarded} time hinges on the distance to point A, $d$. On the contrary, $\rm TeV$ photons are directed toward Earth, causing their {retarded} time to be influenced by projection in this direction, $d \cos \theta$.
}
\label{fig:schem}
\end{figure}
We consider that the emission radius of MeV photons is comparable to that of $\rm TeV$ photons. {A sketch is plotted in Figure \ref{fig:schem}.} In this case, we cannot neglect the impact of the size of the MeV emission region on the collision angles between photons, incorporating the emission radius of MeV photons and their angular influence on photon collision angles.
For simplicity and ease of analysis, we will disregard the angular separation of $\rm TeV$ photons themselves, which means the collision angle remains relatively consistent between $\theta_{\rm T} = 0$ and $\theta_{\rm T} = 1/\Gamma_{\rm T}$.
It is worth noting that MeV photons primarily originate from larger angles, and those emitted from larger radii have a more significant impact on the calculation of optical depth for the greater angles. However, the beaming effect that enters play causes the actual photon angles to be smaller than $\theta_{\rm M,j}$ where
the angle between the radial direction and the photon’s trajectory is also maximized, at $1/\Gamma_{\rm M}$.
As a result, we assume that photons emanate from $\theta_{\rm M}$. Because both $\theta_{\rm M}$ and $\theta$ are smaller than $1/\Gamma_{\rm M}$, which is much less than 1, they can be treated as small quantities.
With this approximation in mind, we arrive at the following.
\begin{equation}
    \theta_{\rm M} \approx \frac{R-R_{\rm M}}{\Gamma_{\rm M}R}
\end{equation}
and
\begin{equation}
    \theta \approx \frac{\theta_{\rm M}R}{R-R_{\rm M}} \approx \frac{R_{\rm M}}{\Gamma_{\rm M}R}.
\end{equation}
It is evident that even as the radius approaches infinity, $\theta_{\rm M}$ remains constrained to $1/\Gamma_{\rm M}$. This observation implies that as long as $\theta_{\rm M}>1/\Gamma_{\rm M}$, the luminosity we derive is {equivalent} to that {coming} from an entire sphere (isotropic) without accounting for angular effects. 
This indicates that we can use the same formula incorporating isotropic luminosity as the previous article \citep{2023arXiv231000631G} to calculate the number density of MeV photons.
Furthermore, the previous neglect of $\theta_{\rm T}$ also suggests that $\theta_{\rm T} = 1/\Gamma_{\rm T}$ does not significantly exceed $\theta$. Given that the Lorentz factor of $\rm TeV$ photons greatly surpasses that of MeV photons, this approximation holds well for scenarios where $R_{\rm T}/R_{\rm M}$ is not excessively large. On the contrary, for cases where $R_{\rm T}/R_{\rm M}$ is substantial, the situation should closely resemble that {explained} in the \cite{2023arXiv231000631G} scenario.
In the early stages of the $\rm TeV$ emission, the $\rm TeV$ emission radius is small, and the size of the MeV emission region determines the collision angle. However, {the $\rm TeV$ emission radius continuously increases with time}, making the size of the $\rm TeV$ emission region even more crucial.

We now move on to considering time delays. In line with our previous article, we can define the {retarded} time to outline the timing of photons on the light curve. Insight garnered from Figure \ref{fig:schem} leads us to infer:
\begin{equation}
    \delta t_{\rm R} = d(1-\cos \theta) \approx \frac{1}{2}(R-R_{M})\theta^2 \approx \frac{1}{2}(R-R_{M})\left(\frac{R_{\rm M}}{\Gamma_{\rm M}R}\right)^2.
    \label{eq:deltatr}
\end{equation}
We can observe that as the radius $R$ approaches infinity, $\delta t_{\rm R}$ tends to zero ($\delta t_{\rm R} \propto 1/R = 0$), which implies that the colliding photons arrive at Earth simultaneously. This outcome aligns with our intuitions and is entirely logical.

Incorporating all the relevant observational parameters into equation (\ref{eq:tau_theta}) and its associated equations, we ultimately arrive at the computation of the average optical depth for a $\rm TeV$ photon at a specific energy and observational time.
\begin{figure}[ht!]
\plotone{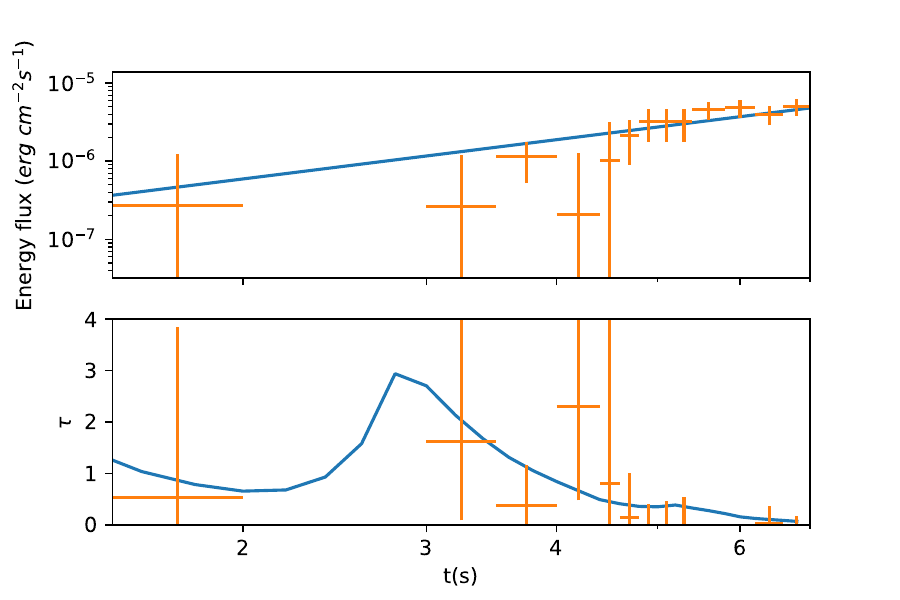}
\caption{{TeV light curve and the optical depth.} The upper panel represents the light curve recorded by \cite{2023Sci...380.1390.} within the initial $7 \ \rm s$, {which includes the energy range} from 0.63 to 1 $\rm TeV$. Data points with central values below 0 have been excluded as they are not suitable for calculating the optical depth. Employing a logarithmic scale, the horizontal axis denotes time, while the vertical axis signifies energy flux. The {orange} points denote the data points, accompanied by the error bars from \cite{2023Sci...380.1390.}. The blue line corresponds to the linear fit undertaken by LHAASO. The blue curve reflects the luminosity of $\rm TeV$ photons before the $\gamma$-$\gamma$ annihilation, leveraging it to determine the optical depth of the data points. This is accomplished by assessing the variance between the intermediate data points and the blue curve, as depicted in the {lower panel}. The {lower panel} delineates the temporal evolution of the optical depth. Utilizing the same horizontal axis as above, the vertical axis now represents optical depth. The blue curve shows the results derived from our model, {with parameters $R_{\rm M}=5.5 \times 10^{15} \rm cm$, $\Gamma_{\rm M} = 100$, $\Gamma_{\rm T} = 400$}.
}
\label{fig:mainresult}
\end{figure}

The lower panel of Figure \ref{fig:mainresult} delineates the relationship between $\bar{\tau}$ and $t_{\rm{T}}$, considering specific parameters for GRB 221009A: $z = 0.151$ \citep{2022GCN.32648....1D}, $E_{\rm T} = 0.81 \rm TeV$ \citep{2023Sci...380.1390.}, $E_{\rm \max,o} = 10$ MeV and $L_{\rm{M}}(t_{\rm{R}})$, $E_{\rm{P}}(t_{\rm{R}})$, $\alpha_{\rm{M}}(t_{\rm{R}})$, $\beta_{\rm{M}}(t_{\rm{R}})$ derived from \cite{2023arXiv230301203A} and \cite{2023ApJ...949L...7F}. During this particular time window, data from the Fermi-GBM (Gamma-ray Burst Monitor) became saturated, thus precluding their utilization \citep{2023arXiv230314172L}. The dataset from \cite{2023ApJ...949L...7F} spans a lengthier time interval, whereas the dataset from \cite{2023arXiv230301203A} offers a finer temporal resolution. Consequently, we employ data from \cite{2023arXiv230301203A} to characterize the pulse and data from \cite{2023ApJ...949L...7F} for other time segments.

In this context, $t_{\rm{T}} = T - T^*$, with $T^* = 226\ \rm s$ \citep{2023Sci...380.1390.}. To achieve better alignment with the data points and meet specific constraints, certain parameter values have been adopted: $R_{\rm M}=5.5 \times 10^{15} \rm cm$, $\Gamma_{\rm M} = 100$, $\Gamma_{\rm T} = 400$. 
Remarkably, by the 4-second mark, our optical depth is already below unity. Notably, at this point, the collision angle between $\rm TeV$ photons with $\theta_{\rm T}=0$ and MeV photons approximates $\theta_{\rm T, j}$, indicating the validity of our approximation. Furthermore, the selected Lorentz factors for the region closely align with those estimated by the afterglow model.
{The light curve can be converted to optical depth, as shown in the lower panel of Figure \ref{fig:mainresult}. We can see, with the parameters above, the optical depth varying with time is well modeled.}

{A complete seeking the parameter space should perform Markov Chain Monte Carlo (MCMC) simulations. Taking into account the complexity} of our model, performing the MCMC simulations incurs a relatively substantial computational expense. For instance, when our sample size reached 100, the process took over an hour to complete. It is important to note that our primary aim is to illustrate the capability of our model in elucidating the flux dip, rather than precisely determining the values of these parameters. {We illustrate the influence of the parameter choosing effect on the optical depth in the Appendix.}


{In Figure \ref{fig:mainresult}, the optical depth curve first decreases, then rises, and finally decreases again. Both declines are due to a decrease in optical depth with increasing distance. The rise is caused by a sudden increase in the flux of $\rm MeV$ photons. Finally, it is worth noting that the decline here is much faster than estimated by the previous method.}

In the preceding approach, the initial collision angle was governed by $\theta_{\rm T}$, which remains constant throughout time. Consequently, it remains unchanged with the initial time. On the contrary, the current method derives the initial collision angle from $\theta_{\rm M}$ and the initial distance, where the latter increases as time progresses. This dynamic causes the initial angle to decrease over time, leading to a more rapid reduction in optical depth with the passage of initial time. Similarly, this phenomenon aligns with observations from the simplified model proposed by \citep{2011ApJ...726L...2Z}:
\begin{equation}
     \tau(\theta_{\rm{T}}) \approx \frac{(1+z)^{2{\alpha}-4}L_{\rm{M}}\epsilon_{\rm{P}}^{\alpha-2} \epsilon_{\rm{T}}^{{\alpha}-1} \sigma_{\rm{T}}R_{\rm{M}}^{2{\alpha}}}{6\times 4^{\alpha_{\rm{M}}}\pi R_{\rm{T}}^{2{\alpha}+1} \Gamma_{\rm M}^{2{\alpha}} m_{\rm{e}}c^3{\alpha}},      
    \label{eq:simple_tua2}
\end{equation}

The optical depth exhibits a power-law decay over time, characterized by $-2{\alpha} -1$. This decay rate is notably faster than the decay rate of the previous method of -1.

Another significant change from the previous method is that our time delay no longer relies on $\Gamma_{\rm T}$. Consequently, there is no longer a trade-off between attaining a larger optical depth and a longer time delay. This progress allows us to simultaneously achieve significant time delay and optical depth.
As a result, we can maintain a low optical depth for the point {before the dip} on the light curve, rendering it observable, while simultaneously ensuring a sufficient optical depth to impede photons within subsequent time intervals.
\begin{figure}[ht!]
\plotone{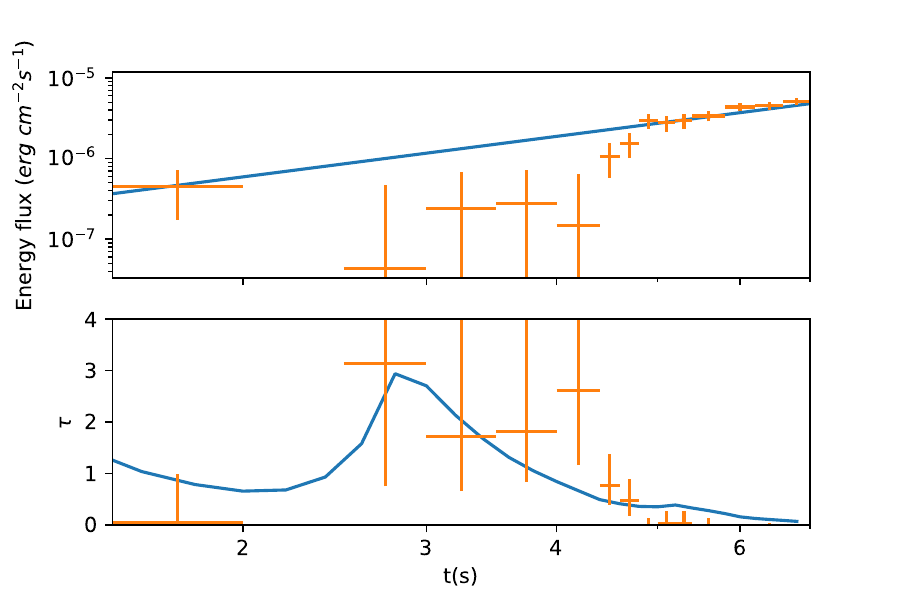}
\caption{{Similar to Figure \ref{fig:mainresult}. The upper panel} displays the light curve data recorded by \cite{2023Sci...380.1390.} spanning the initial $7\  \rm s$, encompassing the energy { range} from 0.3 $\rm TeV$ to infinity. Using a methodology similar to Figure \ref{fig:mainresult}, we generated a comparable plot with the intention of introducing supplementary data points and highlighting the rapid decrease in optical depth.
}
\label{fig:general2}
\end{figure}

We extended the energy range of the photons included in the fit curve from Figure \ref{fig:mainresult} to 0.3 $\rm TeV$ to infinity \citep{2023Sci...380.1390.}, resulting in Figure \ref{fig:general2}.
Due to the broader energy range covered in Figure \ref{fig:general2}, a greater number of data points are available to effectively illustrate fluctuations in optical depth. This plot offers a clearer depiction of the swift reduction in optical depth. The reason behind selecting an energy range of 0.63 to 1 $\rm TeV$ in Figure \ref{fig:mainresult} stems from two key considerations: firstly, this range effectively captures the flux dip in the segmented light curve.
{For lower energy ranges, there is still a significant flux at the bottom of the dip we identified, while at higher energy ranges, we lose the signal before the dip. We believe that lower energy photons are less susceptible to $\gamma$-$\gamma$ annihilation, which results in them not being completely absorbed. On the other hand, higher-energy photons are more prone to undergo $\gamma$-$\gamma$ annihilation, leading to the complete absorption of pre-dip photons.}
{Second}, our choice of $E_{\rm T}$ corresponds to the midpoint of this energy range.

\section{conclusion and discussion}
\label{sec:con}
We have introduced a comprehensive two-zone model that takes into account the angular aspects of MeV photons, offering a robust explanation for the observed dip in the $\rm TeV$ light curve of GRB 221009A. Through the derivation of the collision angle from MeV photon angles and radius, our model successfully introduces a significant time delay between MeV and $\rm TeV$ photons, a crucial factor in allowing the very first $\rm TeV$ photons to evade annihilation with MeV photons. 
{In addition, our results exhibit a peak optical depth exceeding 3, and the optical depth rapidly decreases over time, aligning well with the observed data.}

The performance of our model gives substantial support to our hypothesis that the dip arises from the interaction between MeV and $\rm TeV$ photons. In particular, our model yields constraints on Lorentz factors and emission region dimensions. The success of this endeavor underscores the importance of meticulous geometry considerations when examining MeV-$\rm TeV$ interactions. Subsequent multiwavelength observations with greater detail could provide further validation of the model. {Our result} serves as a compelling illustration of an epoch in which prompt MeV photons genuinely annihilate with $\rm TeV$ photons from the afterglow. It exemplifies a process in which $\rm TeV$ photons transition from an optically thin state (initial phases) to a thicker state (during the dip), and eventually return to an optically thin state (later stages of the light curve).

We assume that MeV photons originate from the widest emission angle possible for interactions with $\rm TeV$ photons. We have two reasons for this assumption. First, a larger emission angle yields a correspondingly larger collision angle and a larger emission area, which in turn determines the optical depth. This could happen once the radius of the MeV photons is close to that of the $\rm TeV$ photons, and the observable region of $\rm TeV$ photons $R_{\rm T}\theta_{\rm T}$ is small as the beaming angle $1/\Gamma_{\rm T}$ is small.
Second, within the studied time frame (2-5 s), the maximum emission angle capable of colliding with $\rm TeV$ photons closely approaches the maximum emission angle observable from Earth. Therefore, this angle fundamentally determines the characteristics of the luminosity curve.

The difference in the optical depth estimated for $\gamma-\gamma$ photon annihilation between this work and \cite{2023arXiv230810477S} may be due to several factors. The most significant factor is that that work considered low-energy photons originating from the afterglow, while this work came from the prompt burst, resulting in a substantial difference in flux. The inferred photon flux reaching Earth from the afterglow hovers around 15 photons per second per square centimeter, whereas the photon flux from the main burst, in the energy range we studied in this work, reaches $1.5 \times 10^3 \, \rm cm^{-2} \, s^{-1}$, differing by two orders of magnitude. 
Additionally, the particle count is directly proportional to the square of the radius, which equivalently translates to the fourth power of the Lorentz factor. The variance in Lorentz factors (spanning 400 to 560) precipitates a radius disparity of approximately one order of magnitude. Finally, this work accounted for energy integration, whereas that work simply multiplied by {the} characteristic energy. Furthermore, the scattering cross section does not attenuate rapidly with energy increase, leading to an approximate one-order-of-magnitude difference.

In our analysis, we have omitted the consideration of the emission angle of $\rm TeV$ photons. Including the emission angle of $\rm TeV$ photons would lead to non-coplanar trajectories for both $\rm TeV$ and MeV photons, significantly increasing the computational complexity of the model. If {we aim} to derive more intricate and precise insights into pertinent parameters, such as the Lorentz factor and emission angles, it might necessitate the execution of Markov Chain Monte Carlo (MCMC) simulations. Such simulations would facilitate the exploration of a wider range of parameter values and enable a more effective estimation of their uncertainties. However, performing such simulations would require significant computational resources.

The results of this study still hold for { gamma-ray bursts with} weaker $\rm TeV$ emissions. The $\rm TeV$ photons from these gamma-ray bursts might originate from different emission radii, leading to varied approximations of the optical depths mentioned in the paper. This variability in the optical depth approximations can impose different constraints on their emission region parameters.

\begin{acknowledgments}
We thank the critical comments from the anonymous reviewer, and for the helpful discussions with A. M. Chen, Kai Wang, Weihua Lei, Jun-Yi Shen, and the hospitality of Yao'an station of Purple Mountain Observatory.
The schematic figure was plotted with GeoGebra.
The English was polished with ChatGPT.
This work is supported by the National Natural Science Foundation of China (grant No. U1931203).
\end{acknowledgments}

\appendix

\section{
The explanation for the parameter selection effects on the optical depth}

{
The optical depth can be determined by three independent parameters, namely the Lorentz factors $\Gamma_T$ and $\Gamma_M$ for TeV and MeV photons, respectively, and the ratio $f_{\rm R}$ of the initial emission radius for MeV and TeV photons (at $t_{\rm R}=1 \rm s$). We adjust these three parameters under the constraints to bring them as close to the data points as possible: the initial radius for TeV photons is greater than that for MeV photons ($f_{\rm R}>1$), and the impact of the incident angle of TeV photons on the results is negligible before the optical depth being less than one.}

\begin{figure}[ht!]
\centering
	\centering
\subfigure[]{\includegraphics[width=.3\textwidth]{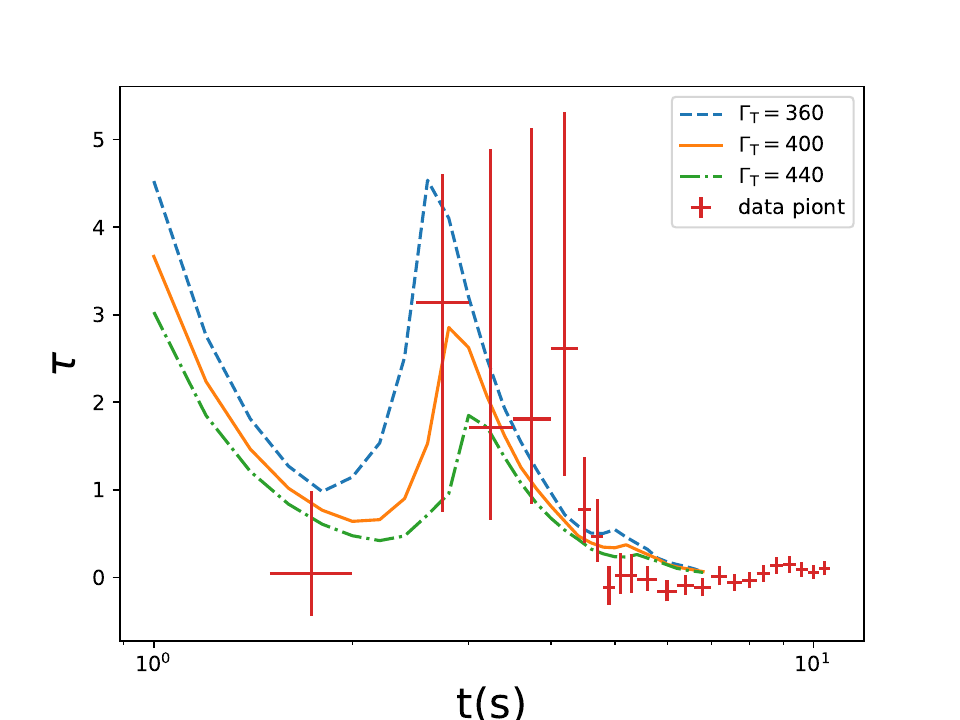}}
\subfigure[]{\includegraphics[width=.3\textwidth]{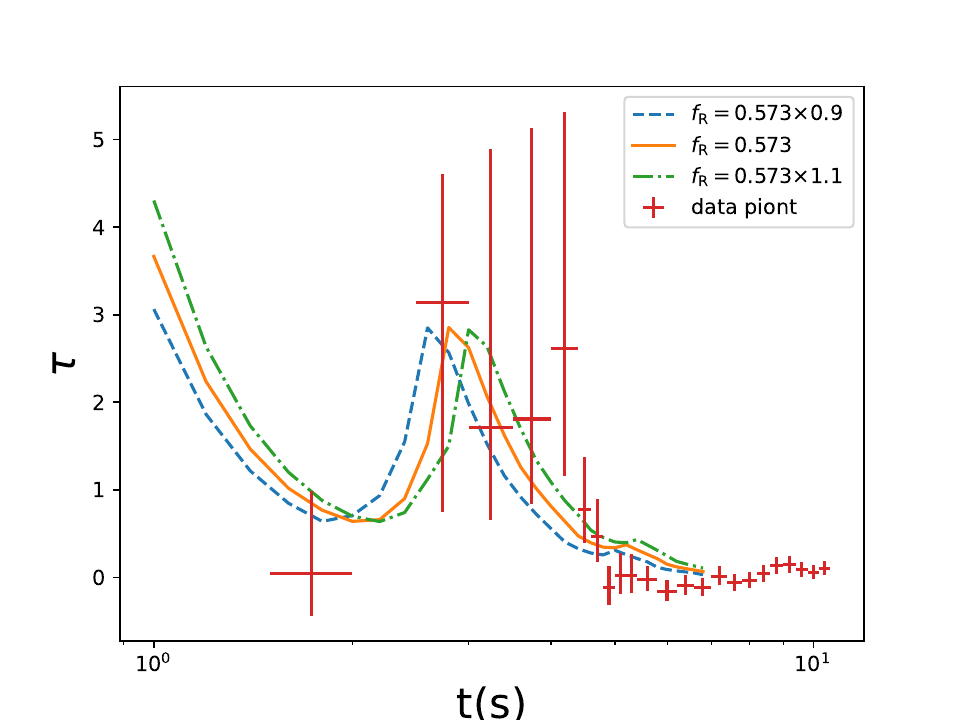}}
\subfigure[]{\includegraphics[width=.3\textwidth]{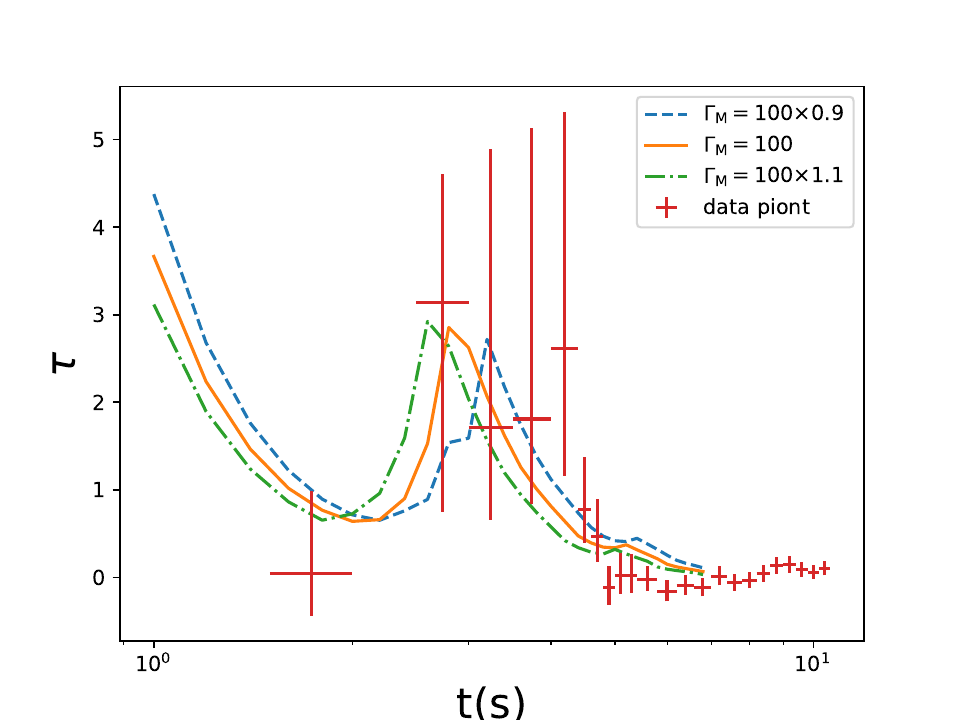}}

\caption{{Illustration on the impact of selecting different parameters on the optical depth. The three subplots on the left (a), center (b), and right (c) respectively demonstrate the influence of changes in the Lorentz factor of the TeV emission region, the ratio of radius between MeV and TeV, and the Lorentz factor of the MeV emission region on the optical depth curve. 
The default values are $\Gamma_{\rm T} = 400$, $f_{\rm R} = 0.573$, $\Gamma_{\rm M} = 100$, respectively.
In each subplot, we show the variation of the optical depth curve when each corresponding parameter is increased or decreased by 10\%.
The original curve is represented by the solid orange line, the curve with parameters decreased by 10\% is shown by the dashed blue line, and the curve with parameters increased by 10\% is depicted by the dashed-dotted green line, respectively.
The red points with error bars in the figure represent the optical depth estimated based on measurements by LHAASO in the energy range from 0.3 to infinity.
It can be observed that the Lorentz factor of the $\rm TeV$ emission region significantly affects the peak value of the optical depth, while the other two parameters mainly influence the peak time.}
}
\label{fig:general4}
\end{figure}

{
We can see the impact of changing parameters on the optical depth in Figure \ref{fig:general4}. 
This figure can be described by two characteristics: the optical depth peak and the peak time. 
Since the starting time of the $\rm TeV$ emission is very close to the peak time of $\rm MeV$ photons, the peak time of the optical depth can be approximated as the time delay $t_{\rm peak} \approx \delta t_{\rm R}$, which is showed in Equation (\ref{eq:deltatr}). 
To establish a connection with our parameters, we use $R_{\rm M} = f_{\rm R} R (\delta t_{\rm R}/1 \rm s)^{-1}$ to replace $R_{\rm M}$ in Equation (\ref{eq:deltatr}). 
At the peak, $R \gg R_{\rm M}$, and $R-R_{\rm M}$ can be approximated as $R$. 
Finally, using the formula $R = 2c\Gamma_{\rm T}t_{\rm peak}$ to replace the remaining $R$ yields:}
\begin{equation}
    t_{\rm peak} \propto \Gamma_{\rm T} f_{\rm R} \Gamma_{\rm M}^{-1}.
\label{tpeak}
\end{equation}
{This roughly explains that the peak time in Figure \ref{fig:general4} increases with the increasing $\Gamma_{\rm T}$ and $f_{\rm R}$, and with decreasing of $\Gamma_{\rm M}$, which can be seen in the timing of the three peaks in Figure \ref{fig:general4}.}

{Next, we can make a similar substitution for Equation (\ref{eq:simple_tua2}) to obtain:}
\begin{equation}
    \tau \propto \Gamma_{\rm M} f_{\rm R}^{-1} \Gamma_{\rm T}^{-2\alpha-3}.
\label{taupeak}
\end{equation}
{This roughly explains that the peak optical depth in Figure \ref{fig:general4} increases with the increasing of $\Gamma_{\rm M}$, and the decreasing of $f_{\rm R}$, and decreases most dramatically with increasing of $\Gamma_{\rm T}$, which can be seen in the $\tau$ values of the three peaks in Figure \ref{fig:general4}.}

\bibliography{TeV-abs-2-zone}{}
\bibliographystyle{aasjournal}


\end{document}